\documentclass[journal]{IEEEtran}

\usepackage{cite}
\usepackage{amsmath}
\usepackage{amssymb}
\usepackage{amscd}
\usepackage{latexsym}
\usepackage{array}

\usepackage{graphicx}

\newcommand{\be}{\begin{equation}}
\newcommand{\ee}{\end{equation}}
\newcommand{\Dlt}{\Delta}
\newcommand{\dlt}{\delta}

\newcommand{\bt}{\beta}
\newcommand{\vp}{\varphi}

\newcommand{\al}{\alpha}
\newcommand{\ra}{\rightarrow}

\newcommand{\gm}{\gamma}

\newcommand{\Gm}{\Gamma}

\newcommand{\lbd}{\lambda}

\newcommand{\cL}{{\cal L}}
\newcommand{\cH}{{\cal H}}

\newcommand{\rgl}{\rangle}
\newcommand{\lgl}{\langle}

\begin{document}

\title{Quantitative Predictions\\ in Quantum Decision Theory}

\author{Vyacheslav I.~Yukalov
        and~Didier~Sornette 

\thanks{V.I. Yukalov and D. Sornette are with the Department of Management,
Technology and Economics, ETH Z\"urich, Swiss Federal Institute of Technology,
Z\"urich CH-8032, Switzerland.}

\thanks{V.I. Yukalov is with the Bogolubov Laboratory of Theoretical Physics,
Joint Institute for Nuclear Research, Dubna 141980, Russia e-mail:
yukalov@theor.jinr.ru.}

\thanks{D. Sornette is with the Swiss Finance Institute, c/o University of Geneva,
40 blvd. Du Pont d'Arve, CH 1211 Geneva 4, Switzerland
e-mail: dsornette@ethz.ch.}

\thanks{Manuscript received ; revised }}


\maketitle

\begin{abstract}
Quantum Decision Theory, advanced earlier by the authors, and illustrated for lotteries
with gains, is generalized to the games containing lotteries with gains as well as
losses. The mathematical structure of the approach is based on the theory of quantum
measurements, which makes this approach relevant both for the description of decision
making of humans and the creation of artificial quantum intelligence. General rules
are formulated allowing for the explicit calculation of quantum probabilities representing
the fraction of decision makers preferring the considered prospects. This provides a
method to quantitatively predict decision-maker choices, including the cases of games
with high uncertainty for which the classical expected utility theory fails. The approach is
applied to experimental results obtained on a set of lottery gambles with gains and losses.
Our predictions, involving no fitting parameters, are in very good agreement with
experimental data. The use of quantum decision making in game theory is described.
A principal scheme of creating quantum artificial intelligence is suggested.
\end{abstract}


\IEEEpeerreviewmaketitle

\section{Introduction}

Classical decision making, based on expected utility theory \cite{Neumann_1}, is known 
to fail in many cases when decisions are made under risk and uncertainty. Numerous 
variants of so-called non-expected utility theories have been suggested to replace 
expected utility theory by using other more complicated functionals. The long list 
of such non-expected utility models can be found in the review articles 
\cite{Machina_2,Machina_3,Bailllon_4}. The non-expected utility theories are, by 
construction, descriptive. By introducing several fitting parameters, such theories 
can be calibrated to some given set of empirical data. However, it is often possible
to have different theories fitting the same set of experiments equally well, so that 
it is difficult to distinguish which of the models is better \cite{Bernstein_5}. 
Moreover, on the basis of such theories, it is impossible to account for the known 
paradoxes arising in classical decision making and to make convincing out-of-sample 
predictions of new sets of empirical data. The non-expected utility theories have 
been thoroughly analyzed in numerous publications confirming the descriptive nature 
of these theories and their inability to perform useful predictions
(see, e.g., \cite{Birnbaum_6,Birnbaum_7,Safra_8,Alnajjar_9,Alnajjar_10}). Thus,
Birnbaum \cite{Birnbaum_6,Birnbaum_7} carefully studied the so-called rank dependent 
utility theory and cumulative prospect theory, concluding that, even with fitting 
parameters, these theories are not able to get rid of paradoxes and moreover create 
new paradoxes. Safra and Segal \cite{Safra_8} state that none of the non-expected utility 
theories can explain all main paradoxes but, on the contrary, distorting the structure 
of expected utility theory, the non-expected utility theories result in several 
non-expected inconsistencies. Al-Najjar and Weinstein \cite{Alnajjar_9,Alnajjar_10} 
present a detailed analysis of non-expected utility theories, coming to the conclusion 
that any variation of expected utility theory "ends up creating more paradoxes and 
inconsistences than it resolves".

The same conclusions apply to the so-called stochastic decision theories
\cite{Hey_11,Ballinger_12,Blavatskyy_13} that are based on underlying deterministic 
theories, decorating them with the probability of making errors in the choice. Introducing 
such probabilities, caused by decision-maker errors, into the log-likelihood functional 
adds several more parameters in the calibration exercise that improve the description 
of the given set of data. But such a stochastic decoration does not change the structure 
of the underlying deterministic theory and does not make predictions possible.

Clearly, the possibility of making predictions can be strongly hindered by the presence 
of unknown or poorly formulated conditions accompanying decision making. For instance, 
there can exist an unknown stochastic environment \cite{Dong_14} or a varying context 
\cite{Andraszewicz_15}. It may also happen that the provided information is imprecise 
and only partially reliable \cite{Aliev_16} or preference relations are incomplete 
\cite{Urena_17} requiring the use of fuzzy logic \cite{Zadeh_64}. In such situations, 
any prediction is likely to be only partial and often merely qualitative.

But even when the posed problem is well defined, suggesting, e.g., a choice between 
explicitly presented lotteries, quantitative predictions as a rule are impossible. In 
particular, the non-expected utility theories mentioned above have been developed exactly 
for such seemingly simple choice between well defined lotteries. And, as is discussed 
above, in many cases, the given lotteries, although being explicitly formulated, contain 
uncertainty not allowing for predictions. It is important to also stress that, in some 
cases of well defined lotteries, predictions based on utility theory are qualitatively 
wrong, as has been demonstrated by Kahneman and Tversky \cite{Kahneman_18}.

In the present paper, we consider the situation when decision making consists in the 
choice between well defined lotteries. We develop an approach allowing for quantitative 
predictions in arbitrary cases, including those where utility theory fails, being unable 
to provide even qualitatively correct conclusions. It is important to emphasize that 
quantitative predictions in our approach can be realized without any fitting parameters. 
So, our approach is not a descriptive, but rather a normative, or prescriptive theory.

Our approach is based on Quantum Decision Theory (QDT), which we developed earlier
\cite{YS_19,YS_20,YS_21,YS_22,YS_23,YS_24,YS_25}. There have been other attempts to apply
quantum techniques to cognitive sciences, as is discussed in the books
\cite{Khrennikov_26,Busemeyer_27,Haven_28,Bagarello_29} and review articles
\cite{YS_30,Agrawal_31,Sornette_32,Ashtiani_33}. However, these attempts were based on
constructing some models for describing particular effects, with the use of several fitting
parameters for each case. Our approach of QDT is essentially different from all those models
in the following facets. First, QDT is formulated as a general theory applicable to any
variant of decision making, but not as a special model for a particular case. Second, the
mathematical structure of QDT is common for both decision theory as well as for quantum
measurements, which has been achieved by generalizing the von Neumann \cite{Neumann_34}
theory of quantum measurements to the treatment of inconclusive measurements and composite
events represented by noncommutative operators \cite{YS_35,YS_36,YS_37,YS_63}.
The third unique feature of QDT is the possibility to develop quantitative predictions 
without any fitting parameters, as has been shown for some simple choices in decision 
making \cite{YS_38}.

The predictions concern the fractions of decision makers choosing the corresponding 
lotteries. In QDT, such fractions are predicted by their corresponding behavioral quantum 
probabilities, as follows from the frequentist interpretation of probabilities and the 
assumption that the population of decision makers are, to a first approximation, 
representative of a homogenous group of individuals making probabilistic choices. The 
scheme for calculating the quantum probabilities is based on our previous demonstration 
that it consists of two terms, called utility and attraction factors. The utility factor 
derives from the utility of each lottery, being defined on prescribed rational grounds. 
The attraction factor represents the irrational side of a choice. The value of the 
attraction factor for a single decision maker and for a given choice is random. However, 
for a society of decision makers, one can derive the {\it quarter law}, which estimates 
the non-informative prior for the absolute value of the average attraction factor as 
equal to $1/4$. In simple cases, the signs of the attraction factors can be prescribed
by the principle of ambiguity aversion. In more complicated situations, a criterion
has been suggested \cite{YS_38} and applied to lotteries with gains.

Here, we extend Ref. \cite{YS_38} by considering lotteries with both gains and losses,
and not just gains. We also improve on the quarter law based on the non-informative prior, 
by including available information on the level of ambiguity characterizing a given set 
of games, thus providing the potential for improved predictions. Moreover, we consider 
the cases for which our previously proposed criterion defining the signs of attraction 
factors does not allow for unique conclusions. We present a generalization of the 
criterion for the sign of the attraction factors that addresses these limitations and 
also applies to lotteries with losses.

The possibility of mathematically formalizing all steps of a decision process, allowing
for quantitative predictions, is important, not merely for decision theory, but also
for the problem of creating an artificial quantum intelligence that could function only
if all operations are explicitly formalized in mathematical terms. We have previously 
mentioned \cite{YS_39} that QDT can provide such a basis for creating artificial quantum
intelligence, since the QDT mathematical foundation is formulated in the same way as
the theory of quantum measurements.

In the present paper, we overcome the limitations of our previous publication
\cite{YS_38} by generalizing QDT along the following directions.

(i) A general method for defining utility factors is advanced, valid for lotteries
with losses as well as for lotteries with gains, or mixed-type lotteries.

(ii) A criterion is formulated for the quantitative classification of attraction
factors for all kinds of lotteries, whether with gains or with losses. In the case
of games with two lotteries, this criterion uniquely prescribes the signs of
attraction factors.

(iii) The quarter law is generalized by taking into account the ambiguity level for
a given set of games. This defines the typical absolute value of the attraction
factor more accurately than the quarter law following from non-informative prior.

(iv) A method for estimating attraction factors for games with multiple lotteries is
described.

(v) The value of our theory is illustrated by comparing its prediction
with empirical results obtained on a set of games containing lotteries with gains
and with losses, for which expected utility theory fails. Our approach results
in quantitative predictions, without fitting parameters, which are in very good 
agreement with empirical data.

(vi) It is shown how the QDT can be applied to game theory. An application is 
illustrated by the prisoner dilemma.

(vii) The general principles for creating artificial quantum intelligence are 
suggested. It is emphasized that artificial intelligence, mimicking the functioning
of human consciousness, should be quantum.

\section{Scheme of Quantum Decision Theory}

In the present section, we briefly sketch the basic scheme of QDT in order to remind
the reader about the definition of quantum probability used in decision theory.
The technical details have been thoroughly expounded in the previous articles
\cite{YS_19,YS_20,YS_21,YS_22,YS_23,YS_24,YS_25}, which allows us to just recall
here the basic notions.

As is mentioned in the Introduction, the mathematical scheme is equally applicable
to quantum decision theory as well as to the theory of quantum measurements
\cite{YS_35,YS_36,YS_37,YS_63}. An event can mean either the result of an estimation
in the process of measurements, or a decision in decision making. In both the cases,
there exist simple events that are operationally testable, that is, clearly observable,
and inconclusive events that are either non-observable or even not well specified.
The typical example in quantum measurements is the double-slit experiment, where the
final registration of a particle by a detector is an operationally testable event,
while the passage through one of the slits is not observable. In decision making, a
straightforward example would be the choice between lotteries under uncertainty.
The final choice of a lottery is an operationally testable event, while the
deliberations on real or imaginary uncertainties in the formulation of the lotteries
or in hesitations of the decision-maker can be treated as inconclusive events.

We consider a set of events $\{ A_n \}$ labelled by an index $n = 1,2,\ldots$. Each
event $A_n$ is put into correspondence with a state $|n \rangle$ of a Hilbert space
$\mathcal{H}_A$, with the family of states $\{ |n \rangle \}$ forming an
orthonormalized basis:
\be
\label{1}
A_n \ra | n \rgl \in \cH_A = {\rm span}\{ | n \rgl \} \;   .
\ee
There also exists another set of events $\{B_\alpha\}$, labelled by an index
$\alpha = 1,2,\ldots$, with each event being in correspondence with a state
$|\alpha \rangle$ of a Hilbert space $\mathcal{H}_B$, the family of the states
$\{ |\alpha \rangle \}$ forming an orthonormalized basis:
\be
\label{2}
 B_\al \ra | \al \rgl \in \cH_B = {\rm span}\{ | \al \rgl \} \;   .
\ee
A pair of events from different sets forms a composite event $A_n \bigotimes B_\alpha$
represented by a tensor-product state $|n \rangle \bigotimes |\alpha \rangle$,
\be
\label{3}
A_n \bigotimes B_\al \ra  | n \rgl  \bigotimes  | \al \rgl  \in  \cH \; ,
\ee
in the Hilbert space
\be
\label{4}
 \cH \equiv \cH_A  \bigotimes \cH_B  =
{\rm span}\{ | n \rgl \bigotimes | \al \rgl \} \;  .
\ee

An event $A_n$ is called {\it operationally testable} if and only if it induces
a projector $|n \rangle \langle n|$ on the space $\mathcal{H}_A$. The event set
$\{ A_n \}$ is assumed to consist of operationally testable events.

A different situation occurs when we have an {\it inconclusive event} being a set
\be
\label{5}
 \mathbb{B} \equiv \{ B_\al , b_\al : \; \al = 1,2, \ldots \}
\ee
of events $B_\alpha$ associated with amplitudes $b_\alpha$ that are random complex
numbers. An inconclusive event corresponds to a state $|B \rangle$ in the space
$\mathcal{H}_B$, such that
\be
\label{6}
\mathbb{B} \ra   | B \rgl = \sum_\al b_\al  | \al \rgl  \in \cH_\mathbb{B} \;  .
\ee
The states $|B_\alpha \rangle$ are not orthonormalized, because of which the
operator $|B \rangle \langle B|$ is not a projector.

A composite event is termed a prospect. Of major interest are the prospects
composed of an operationally testable event and an inconclusive event:
\be
\label{7}
 \pi_n = A_n \bigotimes \mathbb{B} \;  .
\ee
A prospect corresponds to a prospect state in the space $\mathcal{H}$,
\be
\label{8}
 \pi_n \ra |  \pi_n \rgl = | n \rgl \bigotimes | B \rgl  \in \cH \;   ,
\ee
and induces a prospect operator
\be
\label{9}
 \hat P(\pi_n) \equiv |  \pi_n \rgl \lgl \pi_n | \;  .
\ee
The prospect states are not orthonormalized and the prospect operator is not
a projector. The given set of prospects forms a lattice
\be
\label{10}
\cL = \{ \pi_n : \; n = 1,2,\ldots, N_L \} \;   ,
\ee
whose ordering is characterized by prospect probabilities to be defined below.
The assembly of prospect operators $\{ \hat{P}(\pi_n) \}$ composes a positive
operator-valued measure. By its role, this set is analogous to the algebra of 
local observables in quantum theory.

The strategic state of a decision maker in decision theory, or statistical operator
of a system in physics, is a semipositive trace-one operator $\hat{\rho}$ defined
on the space $\mathcal{H}$. The prospect probability is the expectation value
of the prospect operator:
\be
\label{11}
 p(\pi_n) = {\rm Tr} \hat\rho \hat P(\pi_n) \;  ,
\ee
with the trace over the space $\mathcal{H}$. To form a probability measure, the
prospect probabilities are normalized,
\be
\label{12}
 \sum_n p(\pi_n) = 1 \; , \quad 0 \leq p(\pi_n) \leq 1 \;  .
\ee
Taking the trace in (\ref{11}), it is possible to separate out positive-defined
terms from sign-undefined terms, which respectively, are
$$
f(\pi_n)  = \sum_{\al} | \; b_\al\; |^2 \lgl n \al | \hat\rho | n\al \rgl \; ,
$$
\be
\label{13}
 q(\pi_n)  = \sum_{\al\neq\bt} b_\al^* b_\bt \lgl n \al | \hat\rho | n\bt \rgl \; .
\ee
Then the prospect probability reads as
\be
\label{14}
 p(\pi_n) = f(\pi_n) + q(\pi_n) \;  .
\ee
The appearance of a sign-undefined term is typical for quantum theory, describing
the effects of interference and coherence.

Note that the decision-maker strategic state has to be characterized by a statistical 
operator and not just by a wave function since, in real life, any decision maker 
is not an isolated object but a member of a society \cite{YS_37,YS_38}.

An important role in quantum theory is played by the {\it quantum-classical
correspondence principle} \cite{Bohr_40,Bohr_41}, according to which classical
theory has to be a particular case of quantum theory. In the present consideration,
this is to be understood as the reduction of quantum probability to classical
probability under the decaying quantum term:
\be
\label{15}
 p(\pi_n) \ra f(\pi_n) \; , \quad  q(\pi_n) \ra 0 \; .
\ee
In quantum physics, this is also called {\it decoherence}, when quantum measurements
are reduced to classical measurements. The positive-definite term $f(\pi_n)$, playing
the role of classical probability, is to be normalized,
\be
\label{16}
   \sum_n f(\pi_n) = 1 \; , \quad 0 \leq f(\pi_n) \leq 1 \; .
\ee
From conditions (\ref{12}) and (\ref{16}) it follows
\be
\label{17}
\sum_n q(\pi_n) = 0 \; , \quad -1 \leq q(\pi_n) \leq 1 \;   ,
\ee
which is called the {\it alternation law}.

In decision theory, the classical part $f(\pi_n)$ describes the utility of the
prospect $\pi_n$, which is defined on rational grounds. In that sense, a prospect
$\pi_1$ is more useful than $\pi_2$ if and only if
\be
\label{18}
  f(\pi_1) >  f(\pi_2) \quad (more\; useful) \;    .
\ee

The quantum part $q(\pi_n)$ characterizes the attractiveness of the prospect,
which is based on irrational subconscious factors. Hence a prospect $\pi_1$ is more
attractive than $\pi_2$ if and only if
\be
\label{19}
 q(\pi_1) >  q(\pi_2) \quad (more\; attractive ) \;   .
\ee

And the prospect probability (\ref{14}) defines the summary preferability of the
prospect, taking into account both its utility and attractiveness. So, a prospect
$\pi_1$ is preferable to $\pi_2$ if and only if
\be
\label{20}
p(\pi_1) >  p(\pi_2) \quad (preferable) \;   .
\ee

The structure of the quantum probability (\ref{14}), consisting of two parts,
one showing the utility of a prospect and the other characterizing its attractiveness,
is representative of real-life decision making, where both these constituents are 
typically present. Quantum probability, taking into account the rationally defined 
utility as well as such an irrational behavioral feature as attractiveness, can be 
termed as behavioral probability.

It is worth stressing that QDT is an intrinsically probabilistic theory. This is
different from stochastic decision theories, where the choice is assumed to be
deterministic, while randomness arises due to errors in decision making. The
probabilistic nature of QDT is not caused by errors in decision making, but it is
due to the natural state of a decision maker, described by a kind of statistical
operator. Upon the reduction of QDT to a classical decision theory, it reduces
to a probabilistic variant of the latter, since decisions under uncertainty are
necessarily probabilistic \cite{Guo_42}. As mentioned above, the description of 
a decision maker strategic state by a statistical operator, and not by a wave 
function, emphasizes the fact that any decision maker is not an absolutely isolated 
object but rather a member of a society, who is subjected to social interactions 
\cite{YS_37,YS_38,Brock_43}. When comparing theoretical predictions with empirical 
data, it follows from the logical structure of QDT that one has to compare the 
theoretically calculated probability (\ref{14}) with the fraction of decision
makers preferring the considered prospect.

\section{General Definition of Utility Factors}

In this section, we describe the general method for defining utility factors
for a given set of lotteries containing both gains as well as losses.

Let a set of payoffs be given
\be
\label{21}
X_n = \{ x_i : \; i = 1,2,\ldots, N_n \} \; ,
\ee
in which payoffs can represent either gains or losses, being, respectively positive
or negative. The probability distribution over a payoff set is a lottery
\be
\label{22}
 L_n = \{ x_i , p_n(x_i) : \; i = 1,2,\ldots, N_n \} \;  ,
\ee
with the normalization condition
\be
\label{23}
\sum_i p_n(x_i) = 1 \; , \quad 0 \leq p_n(x_i) \leq 1 \;   .
\ee
The lotteries are enumerated by the index $n = 1,2,\ldots,N_L$. Under a utility
function $u(x)$, the expected utility of lottery $L_n$ is
\be
\label{24}
U(L_n) = \sum_i u(x_i) p_n(x_i) \quad ( n = 1,2, \ldots,N_L ) \; .
\ee

Utility functions for gains and losses can be of different signs. Therefore,
the expected utility can also be either positive or negative. When it is negative,
one often uses the notation of the lottery cost
$$
C(L_n) \equiv - U(L_n) = | U(L_n) | \quad (U(L_n) < 0 ) \;   .
$$
An expected utility is positive, when in its payoffs gains prevail. And it is negative,
when losses overwhelm gains.

As has been explained in Ref. \cite{YS_38}, the choice between the given lotteries
in any game is always accompanied by uncertainty related to the decision-maker
hesitations with respect to the formulation of the game rules, understanding of the
problem, and his/her ability to decide what he/she considers the correct choice. 
All these hesitations form an inconclusive event denoted above as $\mathbb{B}$. 
Therefore a choice of a lottery $L_n$ is actually a composite event, or a prospect
\be
\label{25}
 \pi_n = L_n \bigotimes \mathbb{B} \quad  ( n = 1,2, \ldots,N_L ) \; .
\ee
Here we denote the action of a lottery choice and a lottery by the same latter $L_n$,
which should not lead to confusion. The utility factor $f(\pi_n)$ characterizes the
utility of choosing a lottery $L_n$. Since QDT postulates that the choice is probabilistic, 
it is possible to define the average quantity over the set of lotteries,
\be
\label{26}
U = \sum_{n=1}^{N_L} f(\pi_n) U(L_n) \;   ,
\ee
playing the role of a normalization condition for random expected utilities \cite{Gul_44}.

The utility factor represents a classical probability distribution and can be found
from the conditional minimization of Kullback-Leibler information
\cite{Kullback_44,Kullback_45}. The use of the Kullback-Leibler information for
defining such a probability distribution is justified by the Shore-Jonson
theorem \cite{Shore_46} stating that there exists only one distribution satisfying 
consistency conditions, and this distribution is uniquely defined by the minimum of 
the Kullback-Leibler information, under given constraints. The role of the constraints 
here are played by the normalization conditions (\ref{16}) and (\ref{26}). Then the 
information functional reads as
$$
I [\; f\; ] = \sum_{n=1}^{N_L} f(\pi_n) \ln \frac{f(\pi_n)}{f_0(\pi_n)} +
$$
\be
\label{27}
  +
\gm \left [ \sum_{n=1}^{N_L} f(\pi_n) - 1 \right ] +
\bt \left [ U - \sum_{n=1}^{N_L} f(\pi_n) U_n \right ] \;  ,
\ee
where $f_0(\pi_n)$ is a prior distribution, $U_n \equiv U(L_n)$, and $\beta$ and 
$\gamma$ are Lagrange multipliers.

As boundary conditions, it is natural to require that the utility factor of a lottery 
with asymptotically large expected utility would tend to unity,
\be
\label{28}
f(\pi_n) \ra 1 \quad (U_n \ra \infty ) \;   ,
\ee
while the utility factor of a lottery with asymptotically large cost, would go to zero,
\be
\label{29}
f(\pi_n) \ra 0 \quad (U_n \ra -\infty ) \;   .
\ee
Also, the utility factors, as their name implies, have to increase together with
the related expected utilities,
\be
\label{30}
 \frac{\dlt f(\pi_n)}{\dlt U_n} \geq 0 \;  .
\ee
Minimizing the information functional (\ref{27}) results in the utility factors
\be
\label{31}
 f(\pi_n) = \frac{f_0(\pi_n) e^{\bt U_n}}{\sum_n f_0(\pi_n) e^{\bt U_n}} \;  ,
\ee
with a non-negative parameter $\beta$.

If one assumes that the prior distribution is uniform, such that $f_0(\pi_n) = 1/N_L$,
then one comes to the utility factors of the logit form. However, the uniform
distribution does not satisfy the boundary conditions (\ref{28}) to (\ref{29}).
Therefore a more accurate assumption, taking into account the boundary conditions,
should be based on the Luce choice axiom \cite{Luce_47,Luce_48}. According to this
axiom, if an $n$-th object, from the given set of objects, is scaled by a quantity
$\lambda_n$, then the probability of its choice is
\be
\label{32}
f_0(\pi_n) = \frac{\lbd_n}{\sum_n\lbd_n } \; .
\ee

In our case, the considered objects are lotteries and they are scaled by their
expected utilities. So, for the non-negative utilities, we can set
\be
\label{33}
\lbd_n = U_n \quad (U_n \geq 0 ) \;   ,
\ee
while for negative utilities,
\be
\label{34}
  \lbd_n = \frac{1}{| U_n| } \quad (U_n < 0 ) \;  .
\ee
Expression (\ref{34}) is chosen in order to comply with Luce's axiom together
with the ranking of preferences with respect to losses.

Generally, utilities can be measured in some units, say, in monetary units $M$.
Then we could use dimensionless scales $\lambda_n$ defined as $U_n/M$ and $M/U_n$
for gains and losses, respectively. Obviously, expression (\ref{32}) is
invariant with respect to units in which $\lambda_n$ is measured. Therefore, for
simplicity of notation, we assume that utilities are dimensionless.

Thus, the utility factor (\ref{31}), with prior (\ref{32}), is
\be
\label{35}
 f(\pi_n) = \frac{\lbd_n e^{\bt U_n}}{\sum_n \lbd_n e^{\bt U_n}} \quad
(\bt \geq 0 ) \;  .
\ee
In particular, when gains prevail, so that all expected utilities are non-negative,
then
\be
\label{36}
 f(\pi_n) = \frac{U_n e^{\bt U_n}}{\sum_n U_n e^{\bt U_n}} \quad
(\forall \; U_n \geq 0 ) \;  .
\ee
While, when losses prevail, and all expected utilities are negative, then
\be
\label{37}
  f(\pi_n) = \frac{|U_n|^{-1} e^{-\bt |U_n| }}{\sum_n | U_n |^{-1}e^{-\bt| U_n|}}
\quad  (\forall \; U_n < 0 ) \;  .
\ee
In the mixed case, where the utility signs can be both positive and negative,
one has to employ the general form (\ref{35}).

The parameter $\beta$ characterizes the belief of the decision maker with respect 
to whether the problem is correctly posed. Under strong belief, one gets
\begin{eqnarray}
\label{38}
f(\pi_n) = \left \{ \begin{array}{ll}
1 , ~~ & ~~ U_n = \max_n U_n \\
0 , ~~ & ~~ U_n \neq \max_n U_n \end{array}
\quad (\bt \ra \infty) \; , \right.
\end{eqnarray}
which recovers the classical utility theory with the deterministic choice of 
a lottery with the largest expected utility. In the opposite case of weak belief, 
when uncertainty is strong, one has
\be
\label{39}
f(\pi_n) = \frac{\lbd_n}{\sum_n\lbd_n} \quad ( \bt = 0 ) \;   .
\ee

To explicitly illustrate the forms of the utility factors, let us consider
the often met situation of two lotteries under strong uncertainty, thus,
considering the binary prospect lattice
\be
\label{40}
 \cL = \{ \pi_n : \; n = 1,2 \} \quad (\bt = 0 ) \;  ,
\ee
with zero belief parameter. Then, if in both the lotteries gains prevail,
we have
\be
\label{41}
 f(\pi_n) = \frac{U_n}{U_1 + U_2} \quad ( U_1 \geq 0 , \; U_2 \geq 0 ) \;  .
\ee
When losses are prevailing in the two lotteries, then
\be
\label{42}
f(\pi_n) = 1 - \; \frac{ |U_n| }{|U_1| + |U_2|} \quad ( U_1 < 0 , \; U_2 < 0 ) \;    .
\ee
And if one expected utility is positive, say that of the first lottery, while
the other utility is negative, then the utility factor for the first lottery is
\be
\label{43}
f(\pi_1) = \frac{U_1|U_2|}{U_1 | U_2| + 1} \quad ( U_1 > 0 , \; U_2 < 0 ) \;    ,
\ee
respectively, $f(\pi_2) = 1 - f(\pi_1)$.

In this way, the utility factors are explicitly defined for any combination 
of lotteries in the given game, with the payoff sets containing gains as well 
losses.

\section{Classification of Lotteries by Attraction Indices}

We now give a prescription for defining the attraction factors. By its definition, 
an attraction factor quantifies how each of the given lotteries is more or less 
attractive. The attractiveness of a lottery is composed of two factors, possible 
gain and its probability. It is clear that a lottery is more attractive, when 
it suggests a larger gain and/or this gain is more probable. In other words,
a more attractive lottery is more predictable and promises a larger profit.
On the contrary, a lottery suggesting a smaller gain or a larger loss and/or
higher probability of the loss, is less attractive. A less certain lottery is
less attractive, since it is less predictable, which is named as uncertainty
aversion or ambiguity aversion. Below we give an explicit mathematical formulation
of these ideas.

Let us introduce, for a lottery $L_n$, the notation for the {\it minimal gain}
\be
\label{44}
g_n \equiv \min_i \{ x_i \geq 0 : \; x_i \in L_n \}
\ee
and for the {\it minimal loss}
\be
\label{45}
l_n \equiv \max_i \{ x_i \leq 0 : \; x_i \in L_n \}   .
\ee
These quantities characterize possible gains and losses in the given lotteries.

But payoffs are not the only features that attract the attention of decision makers.
In experimental neuroscience, it has been discovered that, during the act of choosing, 
the main and foremost attention of decision makers is directed to the payoff 
probabilities \cite{Kim_49}. We capture this empirical observation by considering 
different weights related to payoffs and to their probabilities in the characterization 
of the lottery attractiveness. Specifically, the weight of a payoff $x$ should be much 
smaller than the weight of its probability $p(x)$. We quantitatively formulate this 
by choosing weights proportional respectively to $x$ for the payoff versus $10^{p(x)}$ 
for its probability. The later term is motivated by the decimal number system.
This leads us to defining the {\it lottery attractiveness}
\be
\label{46}
 a_n \equiv a_n(L_n) \equiv \sum_i x_i 10^{p_n(x_i)} \; .
\ee
And the related relative quantity can be termed the {\it attraction index}
\be
\label{47}
\al_n = \al_n(L_n) \equiv \frac{a_n}{\sum_m | a_m| } \; .
\ee
The latter satisfies the normalization condition
\be
\label{48}
 \sum_n | \al_n| = 1 \;  .
\ee

The notion of the lottery attraction index makes it straightforward to classify all
lotteries from the considered game onto more or less attractive. Thus a lottery $L_1$
is more attractive than $L_2$, hence
\be
\label{49}
 q(\pi_1) > q(\pi_2 ) \;  ,
\ee
when the attraction index of the first lottery is larger than that of the second,
\be
\label{50}
  \al_1 > \al_2 \;  .
\ee

In the marginal case, when $\al_1 = \alpha_2 \geq 0$, the first lottery is more
attractive if the probability of its minimal gain is smaller than that of the second
lottery,
\be
\label{51}
  \al_1 = \al_2 \geq 0 \; , \quad  p(g_1) < p(g_2 ) \;  .
\ee
For short, this will be denoted as $\alpha_1 - \alpha_2 = + 0$. And in the other 
marginal case, where $\al_1 = \alpha_2 < 0$, the first lottery is more attractive 
if the probability of its minimal loss is larger than that of the second,
\be
\label{52}
  \al_1 = \al_2 < 0 \; , \quad  p(l_1) > p(l_2 ) \;  .
\ee
This, for short, will be denoted as $\alpha_1 - \alpha_2 = + 0$.

The criterion allows us to arrange all the given lotteries with respect to the
level of their attractiveness.

For the particular case of a binary prospect lattice (\ref{40}), the alternation
property (\ref{17}) reads as
\be
\label{53}
 q(\pi_1) + q(\pi_2 ) = 0 \;  .
\ee
Therefore the attraction factors have different signs,
\be
\label{54}
 q(\pi_1) = - q(\pi_2 )  \; .
\ee
The sign of each of the attraction factors is prescribed by the sign of the difference
\be
\label{55}
 \Dlt\al  \equiv  \al_1 - \al_2 \;  .
\ee
If $\Delta \alpha$ is positive, then the attraction factor of the first prospect is 
positive and that of the second is negative. On the contrary, if $\Delta \alpha$ is 
negative, then the attraction factor of the first lottery is negative and that of 
the second is positive. In the marginal case, when $\al_1 = \al_2$, we shall use the 
notations accepted above and explained below (Eqs. (\ref{51}) and (\ref{52})): If the 
first lottery is more attractive, we shall write $\Delta \alpha = + 0$, while when 
the second lottery is more attractive, this will be denoted as $\Delta \alpha = - 0$.

\section{Typical Values of Attraction Factors}

The criterion of the previous section allows us to classify all the lotteries of the
considered game onto more or less attractive. But we also need to define the amplitudes
of the attraction factors. According to QDT, these values are probabilistic variables,
characterizing irrational subjective features of each decision maker. For different
subjects, they may be different. They can also be different for the same subject at
different times \cite{Blavatskyy_13}. Different game setups also influence the values
of the attraction factors \cite{Holt_50}. However, for a probabilistic quantity, it
is possible to define its average or typical value.

\subsection{General considerations}

We consider $N_G$ games, enumerated by $k = 1,2,\ldots,N_G$, with $N_L$ lotteries
in each, enumerated by $n = 1,2,\ldots,N_L$. And let the choice be made by a society 
of $N$ decision makers, numbered by $j = 1,2,\ldots,N$. In a $k$-th game, decision 
makers make a choice between $N_L$ prospects $\pi_{nk}$. The typical value of the 
attraction factor is defined as the average
\be
\label{56}
 \overline q  \equiv \frac{1}{N_G} \sum_{k=1}^{N_G} \;
\frac{1}{N_L} \sum_{n=1}^{N_L} \left | \;
\frac{1}{N} \sum_{j=1}^N q_j(\pi_{nk} ) \; \right | \;  .
\ee
Denoting the mean value of the attraction factor for a prospect $\pi_n$, as
\be
\label{57}
 | \; q(\pi_n)\; | \equiv \frac{1}{N_G} \sum_{k=1}^{N_G} \left | \;
\frac{1}{N} \sum_{j=1}^N q_j(\pi_{nk} ) \; \right | \;  ,
\ee
we can write
\be
\label{58}
 \overline q = \frac{1}{N_L} \sum_{n=1}^{N_L} | \; q(\pi_n)\; | \;  .
\ee

For a large value of the product  $N_G N_L N$, the distribution of the attraction 
factors can be characterized by a probability distribution $\varphi(q)$, which, 
in view of property (\ref{17}), is normalized as
\be
\label{59}
\int_{-1}^1 \vp(q) \; dq = 1\;   .
\ee
The average absolute value of the attraction factor can be represented by the integral
\be
\label{60}
 \overline q =  \int_0^1 q\vp(q) \; dq \; .
\ee
This defines the typical value of the attraction factor that characterizes the level
of deviation from rationality in decision making \cite{YS_51}.

If there is no information on the properties and specifics of the given set of
lotteries in the suggested games, then one should resort to a non-informative prior,
assuming a uniform distribution satisfying normalization (\ref{59}), which gives
$\varphi = 1/2$. Substituting the uniform distribution $\varphi = 1/2$ into the
typical value of the attraction factor (\ref{60}) yields $\bar{q} = 0.25$, which was
named the ``quarter law'' in the earlier paper \cite{YS_38}.

However, it is possible to find a more precise typical value $\bar{q}$ by taking
into account the available information on the given lotteries. For example, it is
straightforward to estimate the level of uncertainty of the lottery set.

\subsection{Choice between two prospects}

When choosing between two lotteries with rather differing utilities, the choice looks
quite easy - the lottery with the largest utility is preferred. But when two lotteries
have very close utilities, choosing becomes difficult. The closeness of the lotteries,
corresponding to two prospects $\pi_1$ and $\pi_2$, can be quantified by the
relative difference
\be
\label{61}
  \dlt f(\pi_1,\pi_2) \equiv \frac{2|f(\pi_1)-f(\pi_2)|}{f(\pi_1)+f(\pi_2)} \;
\times 100 \% \; .
\ee
When the choice is between just two prospects, whose utility factors are normalized
according to condition (\ref{16}), hence when $f(\pi_1) + f(\pi_2) = 1$, then the
relative difference simplifies to
\be
\label{62}
 \dlt f = 2|f(\pi_1)-f(\pi_2)| \times 100 \%  \quad ( N_L=2) \;  .
\ee

There have been many discussions concerning choices between similar alternatives 
with close utilities or close probabilities, such that the choice becomes hard to 
make \cite{Thurstone_52,Krantz_53,Rumhelhart_54,Lorentziadis_55}. We refer to such 
situations as ``irresolute''. One of the major problems is how to quantify the 
similarity or closeness of the choices. Several variants of measuring the distance 
between the alternatives $f_1$ and $f_2$ have been suggested, including the linear 
distance $|f_1 - f_2|$, as well as different nonlinear distances $|f_1 - f_2|^m$, 
with $m > 0$. We propose that the value of $\delta f$ that serves as an upper
threshold, below which the lotteries are irresolute, should not depend on the
exponent $m$ used in the definition of the distance. Therefore, in order for the
exponent $m$ not to influence the boundary value, one has to require the invariance
of the distance with respect to the exponent $m$ at the threshold, so that
the critical threshold value should obey the equality: $(\delta f_c)^m = \delta f_c$
for any $m > 0$. The latter reads explicitly as
$$
[\dlt f_c(\pi_1,\pi_2)]^m = \dlt f_c(\pi_1,\pi_2),
$$
where $\dlt f_c(\pi_1,\pi_2)$ is measured in percents. This equation is valid for 
arbitrary $m$ only for $\dlt f_c(\pi_1,\pi_2) = 1\%$. Hence the critical boundary 
value equals $1\%$. Thus the lotteries, for which the {\it irresoluteness criterion}
\be
\label{63}
\dlt f(\pi_1,\pi_2) < 1\%
\ee
is valid, are to be treated as close, or similar, and the choice between them,
as irresolute.

The next question is how the irresoluteness in the choice influences the typical
attraction factor. Suppose that the fraction of irresolute games equals $\nu$.
Then the following properties of the
distribution $\varphi(q)$ over admissible attraction factors should hold.

In the presence of irresolute games $(\nu > 0)$ for which the irresoluteness criterion 
holds true, the probability that the attraction factor is zero is asymptotically 
small,
\be
\label{64}
 \lim_{q\ra 0} \vp(q) = 0 \quad (\nu > 0 ) \;  .
\ee
In other words, this condition means that, on the manifold of all possible games,
absolutely rational games form a set of zero measure.

If not all games are irresolute $(\nu < 1)$, the probability of the maximal
absolute value of the attraction factor is asymptotically small,
\be
\label{65}
 \lim_{|q|\ra 1} \vp(q) = 0 \quad (\nu < 1 ) \;   .
\ee
That is, on the manifold of all possible games, absolutely irrational games
compose a set of zero measure.

Often employed as a prior distribution in standard inference tasks 
\cite{Devroye_56,Mackay_57,Cover_58}, the simplest distribution that obeys the 
two conditions (\ref{64}) and (\ref{65}) is the beta distribution that, under 
normalization (\ref{59}), reads
\be
\label{66}
 \vp(q) = \frac{|q|^\nu ( 1 -|q|)^{1-\nu} }{\Gm(1+\nu) \Gm(2-\nu) } \; .
\ee
Using this distribution, expression (\ref{60}) gives the typical attraction 
factor value
\be
\label{67}
\overline q = \frac{1+\nu}{6} \;   .
\ee
Note that the average of $\overline q$ given by (\ref{67}) over the two boundary 
values $\nu = 0$ and $\nu = 1$ gives 
$$
\frac{1}{2} \; \left({1 \over 6} + {2 \over 6}\right) = {1 \over 4} \; ,
$$
thus recovering the non-informative quarter law.

This expression (\ref{67}) can be used for predicting the results of decision 
making. For example, in the case of a binary prospect lattice, the difference 
in the attraction indices (\ref{55}) defines the signs of the attraction factors,
making it possible to prescribe the attraction factors $\bar{q}$ and $- \bar{q}$ 
to the considered prospects.

\subsection{Choice between more than two prospects}

When there are more than two prospects in the considered game, we propose the 
following procedure to estimate the attraction factors. Using the classification
of the prospects by the attraction indices, as is described in the previous
section, it is straightforward to arrange the prospects in descending order
of attractiveness,
\be
\label{68}
  q(\pi_n) > q(\pi_{n+1} ) \quad ( n= 1,2,\ldots,N_L-1 ) \; .
\ee
Let the maximal attraction factor be denoted as
\be
\label{69}
  q_{max} \equiv q(\pi_1) > 0 \; .
\ee
Given the unknown values of the attraction factors, the non-informative prior 
assumes that they are uniformly distributed and at the same time they must obey 
the ordering constraint (\ref{68}). Then, the joint cumulative distribution of 
the attraction factors is given by
$$
{\rm Pr}[q(\pi_1)< \eta_1,..., q(\pi_{N_L})< \eta_{N_L} | \eta_1 \leq \eta_2 \leq ...
\leq \eta_{N_L}] =
$$
\be
= \int_0^{\eta_1} dx_1  \int_{x_1}^{\eta_2} dx_2  ....  \int_{x_{N_L-1}}^{\eta_{N_L}} dx_{N_L}~,
\ee
where the series $\eta_1 \leq \eta_2 \leq ... \leq \eta_{N_L}$ of inequalities ensure 
the ordering. It is then straightforward to show that the average values of the 
$q(\pi_n)$ are equidistant, i.e. the difference between any two neighboring factors, 
on average, is independent of $n$, so that
\be
 \Dlt \equiv \langle q(\pi_n) \rangle - \langle q(\pi_{n+1}) \rangle = const \qquad
  \;  .
 \label{70}
\ee
Taking their average values as determining their typical values, we omit the symbol 
$\langle . \rangle$ representing the average operator and use the previous equation
to represent the $n$-th attraction factor as
\be
\label{71}
 q(\pi_n) =   q_{max}  - ( n-1)\Dlt \; .
\ee
From the alternation property (\ref{17}), it follows that
\be
\label{72}
  q_{max}  = \frac{N_L-1}{2} \; \Dlt \;  .
\ee
The total number of lotteries $N_L$ can be either even or odd, leading to slightly 
different forms for the following expressions.

And the definition of the typical value (\ref{58}) gives
\begin{eqnarray}
\label{73}
\Dlt = \left \{ \begin{array}{ll}
4\overline q/N_L ~~& ~~ (N_L \; even) \\
\\
4\overline qN_L/(N_L^2-1) ~~& ~~ (N_L \; odd)
\end{array} \right. \; .
\end{eqnarray}
Then the maximal attraction factor (\ref{72}) becomes
\begin{eqnarray}
\label{74}
q_{max} = \left \{ \begin{array}{ll}
2\overline q(N_L-1)/N_L ~~& ~~ (N_L \; even) \\
\\
2\overline qN_L/(N_L+1) ~~& ~~ (N_L \; odd)
\end{array} \right. \; .
\end{eqnarray}
Therefore formula (\ref{71}) yields the expressions for all attraction factors
\begin{eqnarray}
\label{75}
q(\pi_n) = \left \{ \begin{array}{ll}
2\overline q\; \frac{ N_L + 1 - 2n}{N_L}  &   (N_L \; even) \\
\\
2\overline q\; \frac{ N_L(N_L + 1 - 2n)}{N_L^2-1} &  (N_L \; odd)
\end{array} \right. \; .
\end{eqnarray}

Let us denote the set of all attraction factors in the considered game as
$$
Q_{N_L} \equiv \{ q(\pi_n) : \; n = 1,2,\ldots, N_L \} \;   .
$$
If there are only two lotteries, then we have
$$
\Dlt = 2\overline q \; , \quad  q_{max} = \overline q  \quad (N_L = 2) \; ,
$$
and the attraction-factor set is
$$
 Q_2 = \{\overline q ,\; -\overline q \} \;  .
$$
In the case of three lotteries,
$$
\Dlt = \frac{3}{2} \; \overline q \; , \quad  q_{max} = \frac{3}{2} \; \overline q
\quad (N_L = 3) \;  ,
$$
and the attraction factor set reads as
$$
Q_3 = \left \{ \frac{3}{2} \; \overline q ,\; 0 , \; -\frac{3}{2} \;\overline q \right \} \;   .
$$
In that way, all attraction factors can be defined.

\section{Quantitative Predictions in Decision Making}

In order to illustrate how the suggested theory makes it possible to give 
quantitative predictions, without any fitting parameters, let us consider 
the set of experiments performed by Kahneman and Tversky \cite{Kahneman_18}. 
This collection of games, including both gains and losses, is a classical 
example showing the inability of standard utility theory to provide even 
qualitatively correct predictions as a result of the confusion caused by very 
close or coinciding expected utilities. Let us emphasize that the choice of 
these games has been done by Kahneman and Tversky \cite{Kahneman_18} in order 
to stress that standard decision making cannot be applied for these games. 
This is why it is logical to consider the same games and to show that the use 
of QDT does allow us not only to qualitatively explain the correct choice, but 
also that QDT provides quantitative predictions for such difficult cases.

In the set of games described below, each game consists of two lotteries $L_n$,
with $n = 1,2$. The number of decision makers is about $100$.

Recall that, as is explained in Sec. III, the choice between lotteries corresponds 
to the choice between prospects (\ref{25}) including the action of selecting 
a lottery $L_n$ under a set of inconclusive events $\mathbb{B}$ representing 
hesitations and irrational feelings. Therefore the choice, under uncertainty,
between lotteries $L_n$ is equivalent to the choice between prospects $\pi_n$.
The choice under uncertainty for the case of a binary lattice can be
characterized by the utility factors (\ref{41}) to (\ref{43}). We take the
linear utility function, whose convenience is in the independence of the
utility factors from the monetary units used in the lottery payoffs. The
attraction factors are calculated by following the recipes described in
Sec. IV and Sec. V.

We compare the prospect probabilities $p(\pi_n)$, theoretically predicted
by QDT, with the empirically observed fractions \cite{Kahneman_18}
$$
p_{exp}(\pi_n) \equiv \frac{N(\pi_n)}{N}
$$
of the decision makers choosing the prospect $\pi_n$, with respect to the total
number $N$ of decision makers taking part in the experiments.

\subsection{Lotteries with gains}

{\bf Game 1}. The lotteries are
$$
 L_1 = \{ 2.5,0.33 \; |\; 2.4, 0.66 \;| \; 0, 0.01 \}  \; , \quad
 L_2 = \{ 2.4, 1 \} \;  .
$$
For this game, we shall show explicitly the related calculations, while
omitting the intermediate arithmetics in the following cases.

The utilities of these lotteries are
$$
U(L_1) = 2.5 \times 0.33 + 2.4 \times 0.66 + 0 \times 0.01 = 2.409 \; ,
$$
$$
U(L_2) = 2.4 \times 1 = 2.4 \; .
$$
Their sum is
$$
U(L_1) + U(L_2) = 2.409 + 2.4 = 4.809 \; .
$$
The utility factors are close to each other,
$$
 f(\pi_1) = \frac{2.409}{4.809} = 0.501 \; , \quad  f(\pi_2) = \frac{2.4}{4.809} = 0.499 \;  .
$$
For the lottery attractiveness (\ref{46}), we find
$$
 a_1 = 2.5 \times 10^{0.33} + 2.4 \times 10^{0.66} + 0^{0.1} = 16.32 \; ,
$$
$$
a_2 = 2.4 \times 10^1 = 24 \;  ,
$$
which gives
$$
a_1 + a_2 = 16.32 + 24 = 40.32 \; .
$$
The attraction indices (\ref{47}) become
$$
  \al_1 = \frac{16.32}{40.32} = 0.405 \; , \quad \al_2 = \frac{2.4}{40.32} = 0.595 \; .
$$
Then the attraction difference (\ref{55}) is
$$
\Dlt \al = 0.405 - 0.595 = -0.19 \;   .
$$
The negative attraction difference tells us that the first lottery is less 
attractive, $q(\pi_1) < q(\pi_2)$, which suggests that the second lottery 
is preferable, $\pi_1 < \pi_2$. The experimental results confirm this, 
displaying the fractions of decision makers choosing the respective lotteries as
$$
 p_{exp}(\pi_1) = 0.18 \; , \quad  p_{exp}(\pi_2) = 0.82 \;  .
$$
Thus, although the first lottery is more useful, having a larger utility
factor, it is less attractive, which makes it less preferable.

\vskip 2mm

{\bf Game 2}. The lotteries are
$$
L_1 = \{ 2.5,0.33 \; |\; 0, 0.67  \}  \; , \quad
L_2 = \{ 2.4,0.34 \; |\; 0, 0.66  \}  \;  .
$$

The following procedure is the same as in the first game. Calculating
the utility factors
$$
 f(\pi_1) = 0.503 \; , \quad  f(\pi_2) = 0.497 \;  ,
$$
we again see that the lottery utilities are close to each other, so it
is difficult to make the choice. For the lottery attractiveness, we have
$$
a_1 = 16.57 \; , \quad a_2 = 5.25 \;   ,
$$
giving the attraction indices
$$
 \al_1 = 0.759 \; , \quad \al_2 = 0.241 \; ,
$$
and the attraction difference
$$
\Dlt \al = 0.518 \;   .
$$
Now the latter is positive, showing that the first lottery is more attractive,
$q(\pi_1) > q(\pi_2)$, which suggests that the first lottery is preferable,
$\pi_1 > \pi_2$. The experimental data for the related fractions are
$$
p_{exp}(\pi_1) = 0.83 \; , \quad  p_{exp}(\pi_2) = 0.17 \;   ,
$$
in agreement with the expectation that the first lottery is preferable.

\vskip 2mm

{\bf Game 3}. The lotteries are
$$
 L_1 = \{ 4,0.8 \; |\; 0, 0.2  \}  \; , \quad
L_2 = \{3, 1 \}  \;   .
$$
We calculate in the prescribed way the utility factors
$$
  f(\pi_1) = 0.516 \; , \quad  f(\pi_2) = 0.484 \;   ,
$$
lottery attractiveness,
$$
a_1 = 25.24 \; , \quad a_2 = 30 \;   ,
$$
and the attraction indices
$$
\al_1 = 0.457 \; , \quad \al_2 = 0.543 \;    .
$$
The negative attraction difference
$$
\Dlt \al = - 0.086
$$
implies that the first lottery is less attractive, $q(\pi_1) < q(\pi_2)$,
which tells us that the second lottery should be preferable, $\pi_1 < \pi_2$. 
Again this is in agreement with the experimental results
$$
p_{exp}(\pi_1) = 0.2 \; , \quad  p_{exp}(\pi_2) = 0.8 \;   .
$$
The first lottery is less preferable, despite it is more useful, having a larger 
utility factor.

\vskip 2mm

{\bf Game 4}. The lotteries are
$$
 L_1 = \{ 4,0.2 \; |\; 0, 0.8  \}  \; , \quad  L_2 = \{ 3,0.25 \; |\; 0, 0.75  \}  \;  .
$$
Calculating the utility factors
$$
f(\pi_1) = 0.516 \; , \quad  f(\pi_2) = 0.484 \;   ,
$$
lottery attractiveness
$$
 a_1 = 6.34 \; , \quad a_2 = 5.33 \;   ,
$$
and the attraction indices
$$
\al_1 = 0.543 \; , \quad \al_2 = 0.457 \;   ,
$$
we find the positive attraction difference
$$
 \Dlt \al =  0.086 \;  .
$$
Hence the first lottery is more attractive $q(\pi_1) > q(\pi_2)$, which
suggests that the first lottery is preferable, $\pi_1 > \pi_2$. The
experimental data
$$
p_{exp}(\pi_1) = 0.65 \; , \quad  p_{exp}(\pi_2) = 0.35
$$
confirm this expectation.

\vskip 2mm

{\bf Game 5}. The lotteries are
$$
 L_1 = \{ 6,0.45 \; |\; 0, 0.55  \}  \; , \quad  L_2 = \{ 3,0.9 \; |\; 0, 0.1  \}  \;  .
$$
The utility factors
$$
f(\pi_1) = 0.5 \; , \quad  f(\pi_2) = 0.5
$$
turn out to be equal, which makes it impossible to decide in the frame of classical 
decision theory based on expected utilities. Then we calculate the lottery attractiveness
$$
a_1 = 16.91 \; , \quad a_2 = 23.83 \;   ,
$$
and the related attraction indices
$$
\al_1 = 0.415 \; , \quad \al_2 = 0.585 \;   .
$$
The negative attraction difference
$$
\Dlt \al = - 0.17
$$
means that the first lottery is less attractive, $q(\pi_1) < q(\pi_2)$, thence
the second lottery is expected to be preferable, $\pi_1 < \pi_2$. This is
confirmed by the empirical data
$$
p_{exp}(\pi_1) = 0.14 \; , \quad  p_{exp}(\pi_2) = 0.86 \;   .
$$

\vskip 2mm

{\bf Game 6}. The lotteries are
$$
  L_1 = \{ 6,0.001 \; |\; 0, 0.999  \}  \; , \quad  L_2 = \{ 3,0.002 \; |\; 0, 0.998  \}  \;  .
$$
Again their utility factors are equal to each other,
$$
f(\pi_1) = 0.5 \; , \quad  f(\pi_2) = 0.5 \;    .
$$
The lottery attractiveness values
$$
a_1 = 6.01 \; , \quad a_2 = 3.01
$$
yield the attraction indices
$$
\al_1 = 0.666 \; , \quad \al_2 = 0.334 \;   ,
$$
whose positive attraction difference
$$
\Dlt \al = 0.332
$$
implies that the first lottery is more attractive, $q(\pi_1) > q(\pi_2)$,
which suggests that the first lottery should be preferable, $\pi_1 > \pi_2$. 
The experimental results are
$$
p_{exp}(\pi_1) = 0.73 \; , \quad  p_{exp}(\pi_2) = 0.27 \; ,
$$
in agreement with the expectation.

\vskip 2mm

{\bf Game 7}. The lotteries are
$$
 L_1 = \{ 6,0.25 \; |\; 0, 0.75  \}  \; , \quad
L_2 = \{ 4,0.25 \; |\; 2, 0.25 \; | \; 0, 0.5  \}  \;  .
$$
Their equal utility factors,
$$
f(\pi_1) = 0.5 \; , \quad  f(\pi_2) = 0.5 \;    ,
$$
do not allow us to make a choice based on their utility. We calculate
the lottery attractiveness
$$
a_1 = 10.67 \; , \quad a_2 = 10.67
$$
and the attraction indices
$$
 \al_1 = 0.5 \; , \quad \al_2 = 0.5 \;   .
$$
Here the attraction difference is zero, $\Delta \alpha = 0$, with the
attraction indices being positive. Therefore, we resort to criterion
(\ref{51}), for which the minimal gains are $g_1^{min} = g_2^{min} = 0$.
We find that
$$
 p_1(g_1^{min}) = 0.75 > p_2(g_2^{min}) = 0.5 \;  .
$$
According to definitions (\ref{51}) and (\ref{52}), the marginal case,
when $\alpha_1 = \alpha_2$ and $p_1(g_1^{min}) > p_2(g_2^{min})$, is denoted
as $\Dlt \al = - 0$. This proposes that the first lottery is less attractive,
according to the negative sign
$$
 \Dlt \al = - 0 \;  .
$$
Thus we find that $q(\pi_1) < q(\pi_2)$, which suggests that the second
lottery is preferable, $\pi_1 < \pi_2$. The experimental results give
$$
 p_{exp}(\pi_1) = 0.18 \; , \quad  p_{exp}(\pi_2) = 0.82 \;  .
$$

\subsection{Lotteries with losses}

In the previous seven games, the lotteries with gains were considered.
We now turn to lotteries with losses.

\vskip 2mm

{\bf Game 8}. The lotteries are
$$
  L_1 = \{ -4,0.8 \; |\; 0, 0.2  \}  \; , \quad  L_1 = \{ -3, 1  \}  \;  .
$$
Following the same general procedure, we find the utility factors
$$
f(\pi_1) = 0.484 \; , \quad  f(\pi_2) = 0.516 \;    ,
$$
lottery attractiveness
$$
a_1 = -25.24 \; , \quad a_2 = -30 \;    ,
$$
and the attraction indices
$$
\al_1 = -0.457 \; , \quad \al_2 = - 0.543 \;    .
$$
The positive attraction difference
$$
\Dlt \al = 0.086
$$
means that the first lottery is more attractive, $q(\pi_1) > q(\pi_2)$, because
of which, we expect that the first lottery is preferable, $\pi_1 > \pi_2$. The
experiments give
$$
p_{exp}(\pi_1) = 0.92 \; , \quad  p_{exp}(\pi_2) = 0.08 \;   ,
$$
confirming that the first lottery is preferable, although its utility factor
is smaller.

\vskip 2mm

{\bf Game 9}. The lotteries are
$$
 L_1 = \{ -4,0.2 \; |\; 0, 0.8  \}  \; , \quad L_2 = \{ -3,0.25 \; |\; 0, 0.75  \}  \;  .
$$
With the utility factors
$$
f(\pi_1) = 0.484 \; , \quad  f(\pi_2) = 0.516 \;   ,
$$
lottery attractiveness
$$
 a_1 = -6.34 \; , \quad a_2 = -5.33 \;   ,
$$
and the attraction indices
$$
 \al_1 = -0.543 \; , \quad \al_2 = - 0.457 \;  ,
$$
the attraction difference is negative,
$$
\Dlt \al = -0.086 \; .
$$
Thence the first lottery is less attractive, $q(\pi_1) < q(\pi_2)$, and we expect
that the second lottery is preferable, $\pi_1 < \pi_2$. The empirical data are
$$
p_{exp}(\pi_1) = 0.42 \; , \quad  p_{exp}(\pi_2) = 0.58 \;    .
$$

\vskip 2mm

{\bf Game 10}. The lotteries are
$$
  L_1 = \{ -3,0.9 \; |\; 0, 0.1  \}  \; , \quad L_2 = \{ -6,0.45 \; |\; 0, 0.55  \}  \;  .
$$
The utility factors are equal,
$$
f(\pi_1) = 0.5 \; , \quad  f(\pi_2) = 0.5 \;    ,
$$
hence both lotteries are equally useful. But the lottery attractiveness
is different,
$$
 a_1 = -23.83 \; , \quad a_2 = -16.91 \;   ,
$$
yielding the attraction indices
$$
\al_1 = -0.585 \; , \quad \al_2 = - 0.415 \;   .
$$
The negative attraction difference
$$
\Dlt \al = -0.17
$$
signifies that the first lottery is less attractive, $q(\pi_1) < q(\pi_2)$,
which hints that the second lottery is preferable, $\pi_1 < \pi_2$. The
experimental results are
$$
p_{exp}(\pi_1) = 0.08 \; , \quad  p_{exp}(\pi_2) = 0.92\;    .
$$

\vskip 2mm

{\bf Game 11}. The lotteries are
$$
 L_1 = \{ -3,0.002 \; |\; 0, 0.998  \}  \; ,
$$
$$
L_2 = \{ -6,0.001 \; |\; 0, 0.999  \} \;  .
$$
The utility factors are again equal to each other,
$$
f(\pi_1) = 0.5 \; , \quad  f(\pi_2) = 0.5 \;  ,
$$
which makes it impossible to employ the classical utility theory.
But the lottery attractiveness
$$
a_1 = -3.01 \; , \quad a_2 = -59.86
$$ 
and the attraction indices
$$
\al_1 = -0.048 \; , \quad \al_2 = - 0.952
$$
show that the attraction difference is positive,
$$
\Dlt \al = 0.904 \;   .
$$
Therefore the first lottery is more attractive, $q(\pi_1) > q(\pi_2)$,
which suggests that the first lottery is preferable, $\pi_1 > \pi_2$.
The experimental data are
$$
p_{exp}(\pi_1) = 0.7 \; , \quad  p_{exp}(\pi_2) = 0.3 \;  .
$$

\vskip 2mm

{\bf Game 12}. The lotteries are
$$
 L_1 = \{ -1,0.5 \; |\; 0, 0.5  \}  \; , \quad L_2 = \{ -0.5, 1  \}  \;    .
$$
Again the equal utility factors,
$$
 f(\pi_1) = 0.5 \; , \quad  f(\pi_2) = 0.5 \;  ,
$$
do not allow for the choice based on the lottery utilities. But calculating
the lottery attractiveness
$$
a_1 = -3.16 \; , \quad a_2 = -5 \; ,
$$
and the attraction indices
$$
\al_1 = -0.387 \; , \quad \al_2 = - 0.613 \; ,
$$
we see that the attraction difference is positive,
$$
\Dlt \al = 0.226 \;   .
$$
This means that the first lottery is more attractive, $q(\pi_1) > q(\pi_2)$,
thence the first lottery is expected to be preferable, $\pi_1 > \pi_2$. The
empirical results are
$$
p_{exp}(\pi_1) = 0.69 \; , \quad  p_{exp}(\pi_2) = 0.31 \;   .
$$

\vskip 2mm

{\bf Game 13}. The lotteries are
$$
L_1 = \{ -6,0.25 \; |\; 0, 0.75  \}  \; ,
$$
$$
L_2 = \{ -4,0.25 \; | \; -2,0.25 \; | \; 0, 0.5  \}  \;   .
$$
The utility factors are again equal,
$$
f(\pi_1) = 0.5 \; , \quad  f(\pi_2) = 0.5 \;    .
$$
For the lottery attractiveness
$$
 a_1 = -10.67 \; , \quad a_2 = -10.67 \;  ,
$$
the attraction indices are also equal,
$$
 \al_1 = -0.5 \; , \quad \al_2 = - 0.5 \;  .
$$
Getting the zero attraction difference, $\Delta \alpha = 0$, with negative
attraction indices, we have to involve criterion (\ref{52}). The minimal
losses are
$$
 l_1^{min} = l_2^{min} = 0 \;  .
$$
And we find
$$
p_1(0) = 0.75 > p_2(0) = 0.5 \;   .
$$
Consequently, the first lottery is more attractive, which can be denoted as
$$
 \Dlt \al = + 0 \; .
$$
The stronger attractiveness of the first lottery, when $q(\pi_1) > q(\pi_2)$,
suggests that the first lottery should be preferable, $\pi_1 > \pi_2$. The
experimental data are
$$
p_{exp}(\pi_1) = 0.7 \; , \quad  p_{exp}(\pi_2) = 0.3 \;   .
$$

\vskip 2mm

{\bf Game 14}. The lotteries are
$$
 L_1 = \{ -5,0.001 \; |\; 0, 0.999  \}  \; , \quad L_2 = \{ -0.005, 1  \}  \;  .
$$
Although the utility factors are equal,
$$
f(\pi_1) = 0.5 \; , \quad  f(\pi_2) = 0.5 \;  ,
$$
but the lottery attractiveness
$$
a_1 = -5.01 \; , \quad a_2 = -0.05 \;
$$
defines different attraction indices
$$
\al_1 = -0.99 \; , \quad \al_2 = - 0.01 \;   .
$$
The negative attraction difference
$$
\Delta \alpha = -0.98
$$
implies that the first lottery is less attractive, $q(\pi_1) < q(\pi_2)$.
Then the second lottery is expected to be preferable, $\pi_1 < \pi_2$.
The experimental results
$$
p_{exp}(\pi_1) = 0.17 \; , \quad  p_{exp}(\pi_2) = 0.83
$$
confirm this expectation.

\subsection{Empirical test of quantitative predictions of empirical choice frequencies}

\begin{table*}[!t]
\renewcommand{\arraystretch}{1.5}
\caption{The theoretical predictions of QDT are compared with the empirical results
from Ref. \cite{Kahneman_18}.
Here: $f(\pi_1)$ is the
utility factor of the first lottery, $\dlt f$ is the irresoluteness distance (\ref{62}),
$\al(L_1)$ is the attraction index of the first lottery, $p(\pi_1)$ is the predicted
probability of choosing the first lottery, $p_{exp}(\pi_1)$ is the experimentally
observed fraction of decision makers preferring the first lottery, $q_{exp}(\pi_1)$
is the empirically observed attraction factor  for the  first lottery, $\Dlt p$ is
the prediction error.}
\label{Table 1}
\centering
\begin{tabular}{|r|c|c|c|c|c|c|c|c|} \hline
 & $f(\pi_1)$ & $\dlt f\%$ & $\al(L_1)$ & $\Dlt\al$ & $p(\pi_1)$ & $p_{exp}(\pi_1)$ & $q_{exp}(\pi_1)$ & $\Dlt p$
\\ \hline
\hline
1   & 0.501  &  0.4    & 0.405    & -0.19 & 0.23 & 0.18 & -0.32 & 0.05  \\ \hline
2   & 0.503  &  1.2    & 0.759    &  0.52 & 0.78 & 0.83 &  0.33 & 0.05   \\ \hline
3   & 0.516  &  6.4    & 0.457    & -0.09 & 0.24 & 0.20 & -0.32 & 0.04   \\ \hline
4   & 0.516  &  6.4    & 0.543    &  0.09 & 0.79 & 0.65 &  0.13 & 0.14   \\ \hline
5   & 0.5    &  0      & 0.415    & -0.17 & 0.23 & 0.14 & -0.36 & 0.09   \\ \hline
6   & 0.5    &  0      & 0.666    &  0.33 & 0.77 & 0.73 &  0.23 & 0.04   \\ \hline
7   & 0.5    &  0      & 0.5      & -0    & 0.23 & 0.18 & -0.32 & 0.05   \\ \hline
8   & 0.484  &  6.4    & -0.457   &  0.09 & 0.76 & 0.92 &  0.44 & 0.16   \\ \hline
9   & 0.484  &  6.4    & -0.543   & -0.09 & 0.21 & 0.42 & -0.06 & 0.21   \\ \hline
10   & 0.5   &  0      & -0.585   & -0.17 & 0.23 & 0.08 & -0.42 & 0.15   \\ \hline
11   & 0.5   &  0      & -0.048   &  0.90 & 0.77 & 0.70 &  0.20 & 0.07   \\ \hline
12   & 0.5   &  0      & -0.378   &  0.23 & 0.77 & 0.69 &  0.19 & 0.08   \\ \hline
13   & 0.5   &  0      & -0.5     & +0    & 0.77 & 0.70 &  0.20 & 0.07   \\ \hline
14   & 0.5   &  0      & -0.99    & -0.98 & 0.23 & 0.17 & -0.33 & 0.06   \\ \hline
\hline
\end{tabular}
\end{table*}

\vskip 2mm

As is explained at the beginning of Sec. VI, the considered lotteries have been
selected by Kahneman and Tversky \cite{Kahneman_18} in order to demonstrate
the failure of standard utility theory. All these lotteries exhibit close or even 
equal expected utilities, which makes the choice between them difficult or even 
undecided in the frame of utility theory. The majority of the lotteries are 
irresolute in the sense of criterion (\ref{63}). But we show that in the frame 
of QDT, this set of lotteries is treatable.

Among the $14$ considered games, the choice is irresolute in $9$ games according 
to rule (\ref{63}), leading to the fraction $\nu= 9/14$ of irresolute games.
Therefore the typical attraction factor (\ref{67}) is predicted to be
\be
\label{76}
\overline q = 0.274 \quad \left ( \nu = \frac{9}{14} \right ) \;   .
\ee
Then the quantitative predictions for each of the games are determined by
the formulas
$$
p(\pi_1) = f(\pi_1) + \overline q\; {\rm sgn}(\Dlt\al) \; ,
$$
\be
\label{77}
p(\pi_2) = f(\pi_2) - \overline q \; {\rm sgn}(\Dlt\al) \; .
\ee
These predictions are compared with the empirical results from Ref. \cite{Kahneman_18} 
in Table 1. For each game, we show the utility factor $f(\pi_1)$ defined by 
equations (\ref{41}) or (\ref{42}), the utility difference (\ref{62}) providing a 
classification of the game irresoluteness, the attraction index (\ref{47}), the 
attraction difference (\ref{55}), and the predicted probability $p(\pi_1)$, which 
is compared with the experimentally observed fraction $p_{exp}(\pi_1)$ of decision 
makers preferring prospect $\pi_1$. We also report the empirically observed attraction 
factor defined by
$$
q_{exp}(\pi_1) = p_{exp}(\pi_1) - f(\pi_1) \;   ,
$$
and the error of our theoretical prediction, as compared to the empirical data,
$$
\Dlt p = | p(\pi_1) -  p_{exp}(\pi_1) | \;   .
$$
The results for $f(\pi_2)$, $\alpha(L_2)$, $p(\pi_2)$, $p_{exp}(\pi_2)$, and 
$q_{exp}(\pi_2)$ are not shown, since they are straightforwardly connected by the 
appropriate normalization conditions.

The median (resp. average) error is $0.07$ (resp.  $0.09$).
which are within the statistical accuracy of the experiment.

We also compare the predicted typical attraction factor (\ref{76}), with that 
calculated by equation (\ref{57}) using the experimental data. The empirically 
found result is
$$
\overline q_{exp} = 0.275 \; ,
$$
which is in beautiful agreement with the predicted value (\ref{76}).

\section{Quantum Decision Making in Game Theory}

We have shown that Quantum Decision Theory (QDT) provides the basis for an accurate 
description of human decision making. Since decision making is also the basis of 
game theory, it is important to delineate the relation of QDT to game theory.

Quantum game theory, originated by Meyer \cite{Meyer_61} considers game theory from 
the perspective of quantum algorithms. There exist several good reviews on quantum
game theory \cite{Eisert_66,Piotrowski_62,Landsburg_67,Guo_68}). Generally, quantum 
game theory, merging game theory and quantum mechanics, suggests two perspectives,
with the common factor between the two perspectives being quantum information.
In one perspective, players are not assumed to be quantum devices, although their 
conscious processes are described by quantum techniques taking into account the 
dual nature of consciousness, consisting of rational as well as irrational components. 
In this approach, games are treated as representing realistic situations in human 
decision making. From the other side, quantum mechanics as such can be treated as 
a collection of quantum games, which sheds insights into the nature of quantum
algorithms. The discussion of these two perspectives can be found in 
Refs. \cite{Khan_69,Khan_70}. The relation of our Quantum Decision Theory to 
game theory is explained below in more details.

\subsection{Reformulation of games into lottery sets}

In the previous sections, it has been shown how QDT is applied to the choice
between several lotteries. To follow the same way, we need to reformulate
a game into a lottery set, after which we can directly employ the same QDT
decision-making approach. For illustration, let us consider the typical
structure of a two-by-two game. That is, we consider two players, each of
which can accomplish two actions, say $A_1$ and $A_2$. Then there are four
strategies
$$
 s_{mn} = \{ A_m ,\; A_n\} \qquad ( m,n = 1,2 ) \;  ,
$$
where the action $A_m$ corresponds to the first player and $A_n$, to the
second player. The game is characterized by eight payoffs $x_j(s_{mn})$,
with $j = 1,2$. The standard form of a payoff matrix is shown in Table II.

As has been explained above, the use of QDT is senseful only when there exists
uncertainty in decision making. But, when all actions are absolutely certain, being
uniquely defined, the QDT reduces to classical decision making, in agreement with
the quantum-classical correspondence principle (\ref{15}). This implies that QDT
should be applied to mixed games, where actions are taken in the presence of
uncertainty. For this purpose, we introduce a probability measure $\{p_j(A_n)\}$,
with the usual properties
$$
0 \leq p_j(A_n) \leq 1 \; , \qquad p_j(A_1) + p_j(A_2) = 1 \; .
$$
The notation $p_j(A_n)$ defines a probability that the $j$-th player takes the
action $A_n$.

The characteristic feature of games is that the payoffs of one player depend on
the action of the other player, accomplished with the related probability. For
example, the payoff $x_1(s_{mn})$ of the first player, taking an action $A_m$,
is conditioned by the other player action $A_n$, accomplished with the probability
$p_2(A_n)$. Then the mixed game, corresponding to the matrix in Table II, can be
reformulated as the choice between lotteries. Thus, the first player has to choose
between two lotteries
$$
 L_1(A_1) = \{ x_1(s_{11}), p_2(A_1) \; | \; x_1(s_{12}), p_2(A_2) \} \; ,
$$
$$
 L_1(A_2) = \{ x_1(s_{21}), p_2(A_1) \; | \; x_1(s_{22}), p_2(A_2) \} \;  ,
$$
that is, the first player has to decide between the actions $A_1$ and $A_2$,
whose payoffs depend on the actions of the second player. Respectively, the second
player chooses between the lotteries
$$
 L_2(A_1) = \{ x_2(s_{11}), p_1(A_1) \; | \; x_2(s_{21}), p_1(A_2) \} \; ,
$$
$$
 L_2(A_2) = \{ x_2(s_{12}), p_1(A_1) \; | \; x_2(s_{22}), p_1(A_2) \} \;  .
$$

Introducing the payoff utility of a strategy $s_{ij}$ for the $n$-th player,
$$
u_n(s_{ij}) \equiv u_n(x_n(s_{ij})) \;   ,
$$
it is straightforward to define the lottery expected utilities, for the first player,
$$
 U(L_{11}) = u_1(s_{11}) p_2(A_1) + u_1(s_{12}) p_2(A_2) \; ,
$$
$$
U(L_{12}) = u_1(s_{21}) p_2(A_1) + u_1(s_{22}) p_2(A_2) \;  ,
$$
and for the second player
$$
 U(L_{21}) = u_2(s_{11}) p_1(A_1) + u_2(s_{21}) p_1(A_2) \; ,
$$
$$
U(L_{22}) = u_2(s_{12}) p_1(A_1) + u_2(s_{22}) p_1(A_2) \;  ,
$$
where $L_{jn} \equiv L_j(A_n)$.

After the game is reformulated into the approach of choosing between lotteries, 
which is usual for decision theory, it is possible to resort to QDT as we illustrate 
in the next subsection.

\begin{table}[!t]
\renewcommand{\arraystretch}{2}
\caption{General form of a payoff matrix}
\label{Table 2}
\centering
\begin{tabular}{|c|c|c|} \hline
player 1 $\setminus$ player 2 &      $A_1$                & $A_2$       \\ \hline  \hline
    $A_1$           & $x_1(s_{11}),\; x_2(s_{11})$ & $x_1(s_{12}),\; x_2(s_{12})$   \\ \hline
    $A_2$   &       $x_1(s_{21}),\; x_2(s_{21})$ & $x_1(s_{22}),\; x_2(s_{22})$   \\ \hline
\hline
\end{tabular}
\end{table}

\subsection{Quantum decision making in the prisoner dilemma problem}

For concreteness, let us consider the prisoner dilemma that is a canonical example
of a game analyzed in game theory. Numerous other games enjoy the same structure
as the prisoner dilemma. Here the action $A_1$ corresponds to keeping silence, not
betraying the other prisoner, while $A_2$ denotes the action of betraying the other
prisoner.

To apply QDT, we need the variant of the prisoner dilemma, where two players
make decisions under uncertainty, without knowing the choice of the other
agent, as has been treated in Ref. \cite{Kahneman_18}. In that setup,
the $j$-th player, taking an action $A_n$ is not aware of the action accomplished
by the other player. The admissible set of actions of the other player is an
inconclusive event
$$
\mathbb{B} = \{ A_1, \; A_2 \} \;   ,
$$
in agreement with definition (\ref{5}).

In order to avoid excessive notations, let us denote the choice of a lottery
$L_j(A_n)$ by the same symbol. Then the prospect of choosing this lottery by
the $j$-th player, under the unknown choice of the other player, is
\be
\label{79}
 \pi_{jn} = L_j(A_n) \bigotimes \mathbb{B} \; .
\ee
Following the techniques of Sec. II, we get the prospect probabilities
\be
\label{80}
 p(\pi_{jn}) = f(\pi_{jn}) + q(\pi_{jn}) \;  .
\ee

Since we aim at analyzing the prisoner dilemma, we should keep in mind that
this game is symmetric, such that the payoffs of the same strategies are the
same for both players:
$$
x_1(s_{11}) = x_2(s_{11}) \equiv a_1 \; , \qquad
x_1(s_{12}) = x_2(s_{21}) \equiv a_2 \; ,
$$
$$
x_1(s_{21}) = x_2(s_{12}) \equiv a_3 \; , \qquad
x_1(s_{22}) = x_2(s_{22}) \equiv a_4 \; ,
$$
where the values $a_i$ are assumed to be non-negative.

For generality, we may interpret the strategy payoffs as utilities, so that
$$
 u_n(s_{ij}) \; \longrightarrow \; x_n(s_{ij}) \;  .
$$
It is assumed that each player knows nothing about the choice of the other player,
who can take any of the two actions with no informative prior, that is, with 
equal probability $p_j(A_n) = 1/2$. Then the expected utilities of the lotteries 
to be chosen by the players are
$$
U(L_{11})=U(L_{21}) = \frac{1}{2} \; (a_1 + a_2) \; ,
$$
$$
U(L_{12})=U(L_{22}) = \frac{1}{2} \; (a_3 + a_4) \;   .
$$
Because of the symmetry, it is sufficient to consider just one of the players.
Thus, the utility factors for the $j$-th player, according to QDT, are
$$
f(\pi_{j1}) = \frac{a_1+a_2}{a_1+a_2+a_3+a_4} \; ,
$$
$$
f(\pi_{j2}) = \frac{a_3+a_4}{a_1+a_2+a_3+a_4} \;   .
$$

The specific feature of the prisoner dilemma is that the relations between
payoffs are such that
$$
 a_1 < a_3 \; , \qquad a_2 < a_4 \;  .
$$
From this, it immediately follows that
$$
f(\pi_{j1}) < f(\pi_{j2}) \;   ,
$$
hence for each prisoner it is more useful to betray the other. This strategy
is the Nash equilibrium in the classical prisoner dilemma.

However, in QDT, we have to consider not merely the utility, but the whole
prospect probability (\ref{80}). As is explained above, a more useful
prospect may turn out to be not the most attractive. In the prisoner dilemma,
each prisoner takes a decision ``betray or not betray" under the uncertainty
of what has been decided by the other prisoner. According to the principle of
uncertainty and risk aversion \cite{YS_21,YS_22,YS_38}, in the choice under
uncertainty, decision makers are more inclined to be passive, not acting. In
the present case, this means that not betraying is a more attractive action,
hence
$$
 q(\pi_{j1}) > q(\pi_{j2}) \;  .
$$
As is explained in Sec. V, under a complete ignorance of the actions of the
other player, the attraction factors can be evaluated by employing the method
of non-informative priors, yielding the quarter law, telling that the absolute
values of the attraction factors are equal to $1/4$. This tells us that the
prospect probabilities are given by the expressions
$$
 p(\pi_{j1}) = f(\pi_{j1}) + 0.25 \; , \qquad
p(\pi_{j2}) = f(\pi_{j2}) - 0.25 \; .
$$
The first of these is the probability for the prospect of not betraying, while
the second one is the probability for the prospect of betraying.

Generally, the utility factors are influenced by the related payoff strategies.
The typical values of utilities can be found by measuring the fraction of
decision makers, taking the related decisions, when knowing the actions chosen
by the other player, as has been done by Kahneman and Tversky \cite{Kahneman_18},
who found the fractions corresponding to $f(\pi_{j1})=0.1$ and $f(\pi_{j2})=0.9$.
A more detailed discussion of this has been given in Ref. \cite{YS_38}.

Summarizing, we find that QDT predicts the prospect probabilities
$$
 p(\pi_{j1}) = 0.35 \; , \qquad  p(\pi_{j2}) = 0.65 \; .
$$
Thence, despite the fact that the utility of not betraying is very small with 
the utility factor $f(\pi_{j1}) = 0.1$, the probability of not betraying is larger, 
being $p(\pi_{j1}) = 0.35$. This difference is predicted by QDT. The empirical
data, observed by Kahneman and Tversky \cite{Kahneman_18}, when each of the players
decides under uncertainty, are the fractions
$$
p_{exp}(\pi_{j1}) = 0.37 \; , \qquad  p_{exp}(\pi_{j2}) = 0.63 \; ,
$$
which is in remarkable agreement with the prediction of QDT.

\section{Prolegomena to Artificial Quantum Intelligence}

As is emphasized in the Introduction, the developed QDT theory not only provides
a good description of human decision making, allowing for quantitative
predictions, but also can serve as a basic scheme for creating artificial
intelligence \cite{Hutter_65}. Actually, as has been shown above, realistic
human decision making can be interpreted as being characterized by quantum rules.
This is not because humans are quantum objects, but because the process of making
decisions is dual, including both a rational evaluation of utility as well
as subconscious evaluation of attractiveness. The dual nature of the decision
process is effectively taken into account by the mathematics of quantum theory.
Since artificial intelligence is assumed to mimic human mental processes, it has
to function similarly to human decision making. That is, an artificial
intelligence has to necessarily employ a type of quantum decision making. It
is in that sense that an artificial intelligence is to be quantum.

Artificial intelligence is not the same as just a powerful computer, as one
often assumes, but it is a rather different device functioning as a human
brain, hence taking into account the dual nature of decision processes.
A principal scheme of an artificial intelligence, functioning according to
Quantum Decision Theory, is shown in Fig. 1.

\begin{figure*}[!t]
\centering
\includegraphics[width=5in]{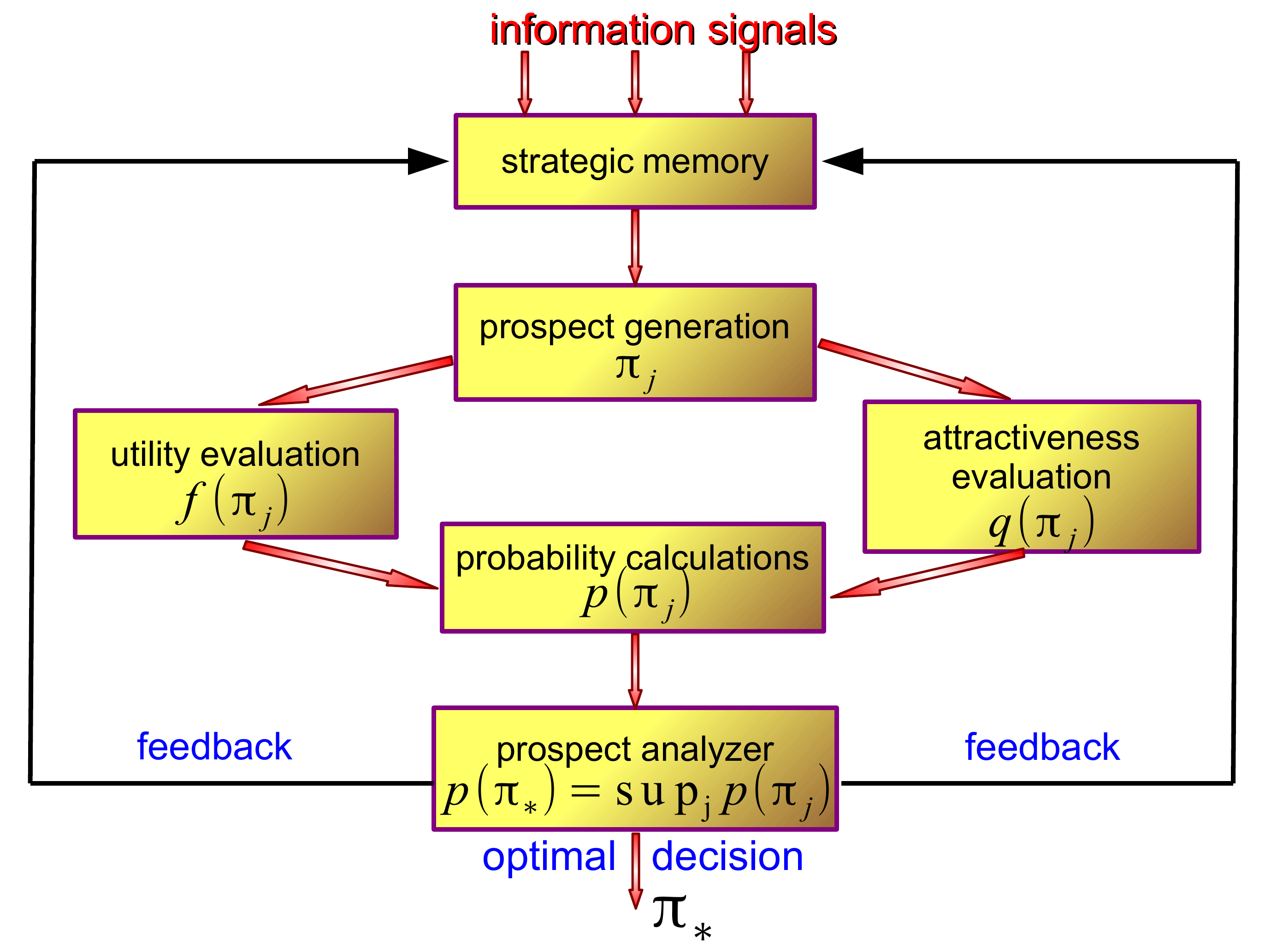}
\caption{Scheme of basic elements of artificial quantum intelligence.}
\label{Fig1}
\end{figure*}

\section{Conclusion}

We have suggested a formulation providing quantitative predictions for the fraction
of decision makers choosing a given prospect among a set of alternatives. This
formulation has the virtue of being parameter-free. The approach is based on
Quantum Decision Theory (QDT) developed earlier by the authors. In the present
paper, the theory has been generalized in several important aspects that are
crucial for the development of such quantitative predictions:

(i) A general method for defining utility factors is advanced, valid for
lotteries with losses as well as for lotteries with gains.

(ii) A criterion is formulated for the quantitative classification of attraction
factors for all kinds of lotteries, whether with gains or with losses. In the
case of games with two lotteries, this criterion uniquely prescribes the signs
of attraction factors.

(iii) The quarter law is generalized by taking into account the irresoluteness
of a given set of games. This defines the typical absolute value of the
attraction factor more accurately than the quarter law based on a non-informative 
prior.

(iv) A method for estimating attraction factors for games with multiple
(more than two) lotteries is described.

(v) The theory is illustrated by a set of games containing lotteries with gains
and with losses, for which expected utility theory fails, while our approach results
in quantitative predictions, without fitting parameters, being in very good agreement
with empirical data.

(vi) It is demonstrated how games considered in game theory can be reformulated 
as sets of lotteries. This opens up the possibility of employing the techniques 
of QDT for mixed games containing uncertainty.

(vii) The mathematical formalization of all steps of decision making provided by
our approach is important not only for accurate quantitative predictions of decision
making by humans, but it can serve as a guiding scheme for creating artificial
intelligence \cite{YS_39}. For this, it is necessary, first of all, to understand 
the basic structure of human decision making. As we have shown, our suggested 
approach captures the main features of decision making by humans, and gives rather 
accurate quantitative predictions. We suggest that the structure of artificial 
intelligence has to include the basic features of the developed QDT approach. 
A principal scheme of artificial quantum intelligence is proposed.

The theory described in the present paper is only a first step in the full
mathematical formalization of the decision making process. Actually, what we have
described is a one-step decision making. In multistep decision making, because
of additional information acquired by agents, the choice can change \cite{Charness_60}.
To characterize the sequence of decisions taken by decision makers who are the members
of a society, it is necessary to generalize the approach by taking into account
dynamical effects influencing the temporal evolution of decisions due to the
exchange of information among social decision makers. First attempts of generalizing
QDT to account for temporal effects, caused by the amount of information among
social decision makers, have been considered in Refs. \cite{YS_25,YS_38}, where it was
assumed that all decision makers in a society simultaneously get the same information.
A more realistic situation of decision makers exchanging information with each other
and varying their decisions accordingly will be presented in a following paper.

\section*{Acknowledgment}

The authors are grateful for discussions to E.P. Yukalova. Financial support
from the ETH Z\"{u}rich Risk Center is appreciated.


\begin{IEEEbiography}[{\includegraphics[width=1in,height=1.25in,clip,keepaspectratio]{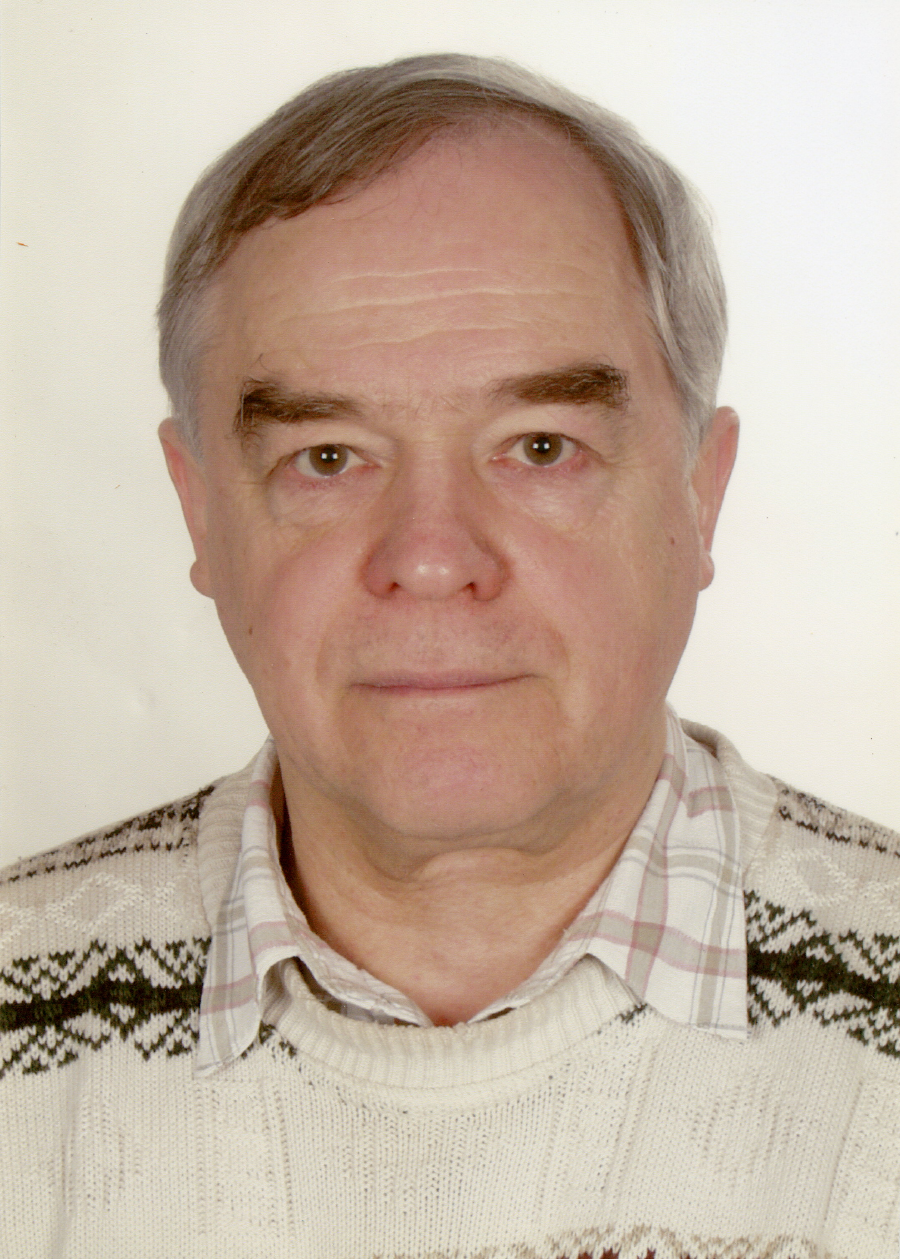}}]{Vyacheslav Yukalov}
received the M.Sc. degree in theoretical physics and
Ph.D. degree in theoretical and mathematical physics from the Physics Faculty
of the Moscow State University, Moscow, Russia, in 1970 and 1974, respectively.
He has also received the Dr.Hab. degree in theoretical physics from the University
of Poznan, Poznan, Poland, and the Dr.Sci. degree in physics and mathematics, from
the Higher Attestation Committee of Russia. He was a Graduate Assistant at the Moscow
State University from 1970 to 1973, before becoming an Assistant Professor,
Senior Lecturer, and an Associate Professor at the Moscow Engineering Physics
Institute, Moscow, Russia, from 1973 to 1984. Since 1984, he has been a Senior
Scientist and Department Head at the Joint Institute for Nuclear Research in Dubna,
Russia, where he currently holds the position of a Leading Scientist. He has authored
about 450 papers in refereed journals and four books. He is also an Editor of five
books and special issues. His current research interests include decision theory,
quantum theory, dynamical theory, complex systems, nonlinear and coherent phenomena,
self-organization, and development of mathematical methods. Prof. Yukalov is a member
of the American Physical Society, American Mathematical Society, European Physical
Society, International Association of Mathematical Physics, and the Oxford
University Society. His scientific awards include the Research Fellowship of the
British Council, Great Britain, from 1980 to 1981, Senior Fellowship of the Western
University, Canada, in 1988, First Prize of the Joint Institute for Nuclear Research,
Russia, in 2000, Science Prize of the Academic Publishing Company, Russia, in 2002,
Senior Fellowship of the German Academic Exchange Program, Germany, in 2003, the
Mercator Professorship of the German Research Foundation, Germany, from 2004 to 2005,
and the Risk-Center Professorship of ETH Z\"{u}rich, in 2015-2016.
\end{IEEEbiography}

\begin{IEEEbiography}[{\includegraphics[width=1in,height=1.25in,clip,keepaspectratio]{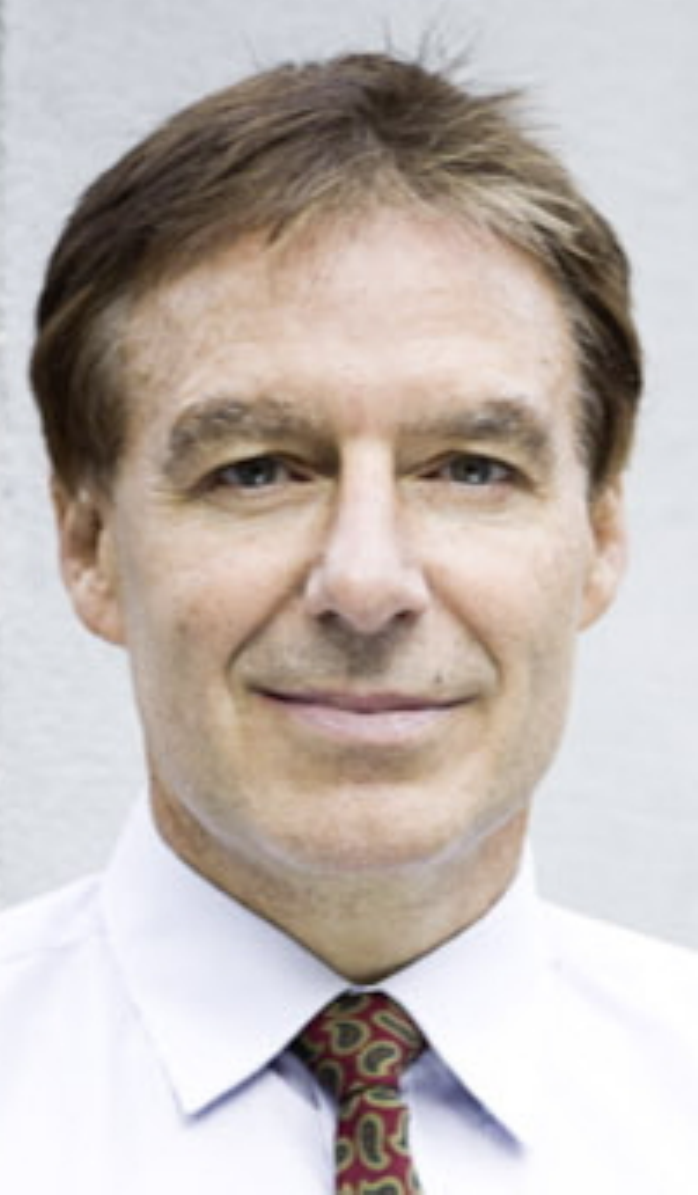}}]{Didier Sornette}
is professor of Entrepreneurial Risks
in the department of Management, Technology and Economics at the Swiss
Federal Institute of Technology (ETH Zurich), a professor
of finance at the Swiss Finance Institute, and is associate member
of the department of Physics and of the department of Earth Sciences at ETH Zurich.
He uses rigorous data-driven mathematical statistical analysis
combined with nonlinear multi-variable dynamical models including
positive and negative feedbacks to study the predictability and control
of crises and extreme events in complex systems, with applications to
financial bubbles and crashes, earthquake physics and geophysics, the
dynamics of success on social networks and the complex system approach
to medicine (immune system, epilepsy and so on) towards the diagnostic of
systemic instabilities. He directs the Financial Crisis Observatory that
now operationally diagnoses financial bubbles in real-time ex-ante
worldwide through a daily scanning of more than 25000 assets.
\end{IEEEbiography}





\end{document}